\documentclass[12pt]{article}
\usepackage{arxiv}
\usepackage{amsmath}
\usepackage{comment}
\usepackage{graphicx}
\usepackage{natbib}

\def\bSig\mathbf{\Sigma}

\begin{document}
\title{TITE-Safety: Time-to-event Safety Monitoring for Clinical Trials}

\author{Michael J. Martens, Qinghua Lian, Brent R. Logan
\\ Division of Biostatistics, Medical College of Wisconsin, Milwaukee, Wisconsin, U.S.A.
\\ Center for International Blood and Marrow Transplant Research, Milwaukee, Wisconsin, U.S.A.}

\maketitle

\label{firstpage}

\begin{abstract}
Safety evaluation is an essential component of clinical trials. To protect study participants, these studies often implement safety stopping rules that will halt the trial if an excessive number of toxicity events occur. Existing safety monitoring methods often treat these events as binary outcomes. A strategy that instead handles these as time-to-event endpoints can offer higher power and a reduced time to signal of excess risk, but must manage additional complexities including censoring and competing risks. We propose the TITE-Safety approach for safety monitoring, which incorporates time-to-event information while handling censored observations and competing risks appropriately. This strategy is applied to develop stopping rules using score tests, Bayesian beta-extended binomial models, and sequential probability ratio tests. The operating characteristics of these methods are studied via simulation for common phase 2 and 3 trial scenarios. Across simulation settings, the proposed techniques offer reductions in expected toxicities of 20\% or more compared to binary data methods and maintain the type I error rate near the nominal level across various event time distributions. These methods are demonstrated through a redesign of the safety monitoring scheme for BMT CTN 0601, a single arm, phase 2 trial that evaluated bone marrow transplant as treatment for severe sickle cell disease. Our R package ``stoppingrule” offers functions to construct and evaluate these stopping rules, providing valuable tools for trial design to investigators.

\end{abstract}

\maketitle

\section{Introduction}
\label{s:intro}

The objectives of phase 2 and 3 clinical trials are to evaluate the efficacy and safety of novel therapies. Safety evaluation is an essential component of these studies; sufficient data need to be collected to convince patients, physicians, and regulatory agencies that the investigational therapy does not pose undue health risks in order for it to gain acceptance. Moreover, the safety profile of a therapy is often not well understood when these trials are designed. To protect the safety of study participants, data on adverse reactions are reviewed periodically by a data and safety monitoring board on an annual or semi-annual basis. Additionally, safety stopping rules are often included to monitor occurrences of specific toxicities of interest such that the trial will be paused if the frequencies of these events exceed expected levels in order to exercise vigilance in safety monitoring throughout the study. 

Effective stopping rules need to strike a balance between providing a high chance of identifying excess safety risks and ensuring the probability of a false positive safety signal is small. Additional challenges entailed in safety monitoring include the repeated assessments of toxicity data throughout the trial and the potentially small numbers of patients evaluated at these interim looks. Statistical procedures have been devised to achieve this balance in yielding both a high power and low type I error rate for detecting elevated toxicity risks under repeated examinations over the course of the trial.

The majority of existing safety monitoring methods treat the toxicity data as binary outcomes that indicate whether a toxicity occurred within a specified evaluation window, for instance, the occurrence of graft rejection within 100 days of a bone marrow transplant. Using a frequentist approach,
\cite{IvanQaqi05} developed versions of the Pocock and O'Brien-Fleming designs \citep{Poco77, OBriFlem79} for sequentially testing binary toxicity data for evidence of excess risk. \cite{Gold87} proposed a truncated sequential probability ratio test (SPRT) that compares a null, expected toxicity rate to a prespecified excessive rate using repeated likelihood ratio tests. A similar strategy by \cite{KullDavi11} employs a maximized SPRT that uses a generalized likelihood ratio which compares the null rate to all elevated toxicity rates. Bayesian beta-binomial models have also been applied to safety monitoring by \cite{GellFoll03} and \cite{ChenChal06}, where a safety signal is flagged if the posterior probability that the toxicity rate exceeds the null rate is sufficiently high.

Because the precise dates of toxicity events are often collected, they can also be treated as time-to-event outcomes. Therefore, similar to interim monitoring of efficacy endpoints, these safety events can be analyzed by time-to-event methods that may offer benefits over binary data strategies including higher power and/or a lower time to signal. Time-to-event based approaches to safety monitoring in clinical trials are scarce, however, particularly when competing risks are at play. For post-marketing surveillance studies, a maximized SPRT \citep{KullDavi11} and alpha spending tests \citep{Silv18, SilvMaro21} were proposed to assess the failure rate, assuming that the event count follows a Poisson process over time. \cite{MartLian25} applied score tests, Bayesian models, and SPRTs to monitor the failure rate in clinical trials under this Poisson process assumption. But, these works do not account for competing risks and are sensitive to violations of the Poisson process assumption. A group sequential test of a cumulative incidence probability can be used for monitoring toxicities in the presence of competing risks \citep{LogaZhan13}, but relies on large sample properties for type I error control and may not provide this control when small numbers of patients are evaluated. Thus, new methods are needed for the  clinical trial setting that can exploit event time information, maintain false positive rate control under frequent or continuous monitoring, and account for competing risks when monitoring safety.

Techniques for sequential evaluation of events that are subject to competing risks have been studied in the context of phase 1 dose-finding trials. These studies perform repeated evaluations of the cumulative incidence of dose limiting toxicities (DLTs) over time in order to allocate doses for newly enrolled patients. For patients who have experienced a DLT or a competing risk within the evaluation period, their DLT status is known. To incorporate data from event-free patients who have not completed evaluation and thus have an undetermined DLT status, \cite{CheuChap00} proposed the TITE-CRM design, an adaptation of the continual reassessment method (CRM) \citep{OQuiPepe90} that includes these patients in the likelihood by using the fractions of the evaluation window they have completed. \cite{YuanLin18} and \cite{LinYuan20} applied this strategy to adapt the Bayesian optimal interval design (BOIN) \citep{LiuYuan15}, modified toxicity probability interval design (mTPI) \citep{JiLiu10}, and keyboard design \citep{YanMand17} to utilize these incomplete observations as well. But, this scheme have not been investigated for safety monitoring to date.

This article adapts this strategy for the evaluation of toxicity risk in clinical trials, an approach termed TITE-Safety. Score tests, Bayesian models, and SPRTs are developed under this approach to construct safety stopping rules for time-to-event safety outcomes while handling competing risks appropriately. An extensive simulation study assesses and compares the operating characteristics of these novel stopping rules to each other, to existing binary data methods, and to techniques that rely on a Poisson process assumption for the cumulative event counts over time. The proposed methods are applied in a redesign of the safety monitoring scheme of a clinical trial that demonstrates how functions from our R package ``stoppingrule" can construct and evaluate these stopping rules. Lastly, we provide guidance for choosing the stopping rule method when designing a trial and discuss future research interests in safety monitoring methodology.

\section{Methods}
\label{s:methods}

\subsection{General Approach}
\label{subsec:m_approach}
Consider a clinical trial where the cumulative incidence of a time-to-event safety outcome at the follow-up time $\tau$ will be monitored for evidence that it exceeds a tolerable level of risk. This surveillance may be performed on the trial population of a single arm, phase 2 trial or within a treatment arm or cohort of a phase 3 trial. The methodology that follows can accommodate time-to-event data with and without competing risks. Without loss of generality, events are attributed to two possible causes: cause 1 events correspond to the toxicity being monitored, while cause 2 events are comprised of the competing risk(s) for toxicity. Let $T$ denote the event time, $\epsilon$ the event cause, $E$ the calendar time of enrollment, and $s$ the calendar time of analysis. Let $F_1(t) = P(T \le t, \epsilon = 1)$ denote the cumulative incidence function for event type 1 and $p = F_1(\tau)$. The monitoring objective can be formalized as a repeated test of the null hypothesis $H_0: p = p_0$ versus the alternative $H_A: p > p_0$, where $p_0$ is the maximum tolerable toxicity rate.

Assume that this monitoring proceeds until a maximum of $N$ patients have been evaluated and let $n(s)$ denote the number of patients who have enrolled by analysis time $s$. Let $\{T_j,\epsilon_j,E_j\}_{j=1}^N$ denote the data vectors from these patients, which are assumed to be independently and identically distributed. At the time $s$, the observed toxicity data for patient $j$ is comprised by the observation time $X_j(s) = \max\{\min\{T_j, \tau, s-E_j\},0\}$ and event cause indicator $\delta_j(s) = I[T_j \le \min\{\tau, s-E_j\}] \cdot \epsilon_j$, where $\delta_j(s) = \epsilon_j$ if the patient had an event and 0 if their event time is censored.

In the absence of censoring, the observations $y_j = I(T_j \le \tau, \epsilon_j = 1)$ are Bernoulli distributed with event probability $p$ and so sequential methods for binary data can be applied to test $H_0$. Likelihood-based approaches may be used for this purpose, including binary Wang-Tsiatis tests \citep{IvanQaqi05,MartLoga23}, Bayesian beta-binomial models \citep{GellFoll03}, and SPRTs \citep{Gold87,KullDavi11}. Under staggered enrollment, however, patient $j$ may be censored at a given analysis, precluding the observation of $y_j$. To incorporate data from such patients into safety evaluation, we utilize an approximate likelihood that permits the extension of these binary monitoring methods to include censored observations. For a patient $j$ enrolled before analysis time $s$, three possibilities exist: (i) $\delta_j(s)=1:$ they experienced a cause 1 event by time $\tau$; (ii) $\delta_j(s)=2:$ they had a cause 2 event by time $\tau$; and (iii) $\delta_j(s)=0:$ they are event-free and have not completed $\tau$ units of follow-up. For cases 1 and 2, $y_j$ is observed and the respective contributions to the likelihood are $p$ and $1-p$. For case 3, we only know they did not have a cause 1 event through $s-E_j$, so their observation is the binary outcome $I(T_j \le s-E_j, \epsilon_j = 1) = 0$; thus, their likelihood contribution is $1-P(T_j \le s-E_j, \epsilon_j=1) =  [1 - w_j(s) p],$
where $w_j(s) = G(s-E_j) = F_1(s-E_j) / F_1(\tau)$ is the cumulative distribution function of cause 1 event times truncated to $[0,\tau]$. Thus, the likelihood at analysis time $s$ is
\begin{equation}
\label{eq:bin_lik}
\begin{alignedat}{2}
L(p,s) &= \prod_{j=1}^{n(s)} p^{I[\delta_j(s)=1]} (1-p)^{I[\delta_j(s)=2]} [1 - w_j(s) p]^{I[\delta_j(s) = 0]} \\
 &= p^{D_1(s)} (1 - p)^{D_2(s)} \prod_{j=1}^{n(s)} [1 - w_j(s) p]^{I[\delta_j(s) = 0]},
\end{alignedat}
\end{equation}
where $D_k(s) = \sum_{j=1}^{n(s)} I[\delta_j(s)=k]$ is the number of patients with cause $k$ events through calendar time $s$. \cite{CheuChap00} used this likelihood structure to extend the CRM dose-finding design to incorporate partial follow-up on patients; this extension is called the TITE-CRM design, where TITE denotes time-to-event. 

The likelihood in (\ref{eq:bin_lik}) has a non-standard form, complicating inference procedures. To obtain a more useful likelihood for inference in dose-finding designs, \cite{LinYuan20} proposed the application of the first-order Taylor approximation $(1-p)^{w_j} \approx 1 - w_j p$ to approximate (\ref{eq:bin_lik}) by an ``extended binomial" likelihood defined as
\begin{align}
\label{eq:ebin_lik}
\widetilde{L}(p,s) &= p^{D_1(s)} (1 - p)^{D_2(s)} \prod_{j=1}^{n(s)} (1 - p)^{w_j(s) I[\delta_j(s) = 0]} = p^{D_1(s)} (1 - p)^{\widetilde{n}(s) - D_1(s)}
\end{align}
where $\widetilde{n}(s) = D_1(s) + D_2(s) + \sum_{j=1}^{n(s)} w_j(s) I[\delta_j(s) = 0]$ represents the ``effective sample size" among patients enrolled at analysis time $s$. Each censored patient contributes a fraction $w_j(s)$ to the sample size, while those who had an event contribute a full unit. This extended binomial likelihood resembles the usual binary likelihood, only differing in that the numbers of trials $\widetilde{n}(s)$ and nonevents $\widetilde{n}(s) - D_1(s)$ can take noninteger values. \cite{LinYuan20} used (\ref{eq:ebin_lik}) to extend the BOIN \citep{LiuYuan15}, mTPI \citep{JiLiu10}, and Keyboard designs \citep{YanMand17} to incorporate censored observations, termed TITE-BOIN, TITE-mTPI, and TITE-Keyboard. To calculate the weights $w_j$, the distribution $G$ must be specified. \cite{CheuChap00} and \cite{LinYuan20} recommended assuming a Uniform$(0,\tau)$ distribution for $G$ such that the weights $w_j(s)$ can be easily computed as $(s-E_j)/\tau$. Simulation studies by these authors showed that this choice work well in various dose finding designs, even for cases where $G$ deviates greatly from a uniform distribution.

In the same spirit, the remainder of this section proposes monitoring methods, which we call TITE-Safety monitoring designs, by using this extended binomial likelihood to construct safety monitoring procedures in a similar manner as for binary toxicity data by assuming that $G$ follows a Uniform distribution. Each sequential test has a decision rule that rejects $H_0$ at analysis time $s$ if $D_1(s) \ge b\big(\widetilde{n}(s)\big),$ where the test's stopping boundary function $b$ is  continuous and monotonically increasing. If all patients complete evaluation without rejecting $H_0$, $H_0$ is accepted and the monitoring terminates.

\subsection{Wang-Tsiatis Test}
\label{subsec:m_wt}
Using a score test based on (\ref{eq:ebin_lik}), the family of Wang-Tsiatis tests for binary data \citep{MartLoga23} are extended to accommodate censored observations. The respective score function and expected information at analysis time $s$ are
\begin{align*}
S(p,s) &= \frac{\partial \log \widetilde{L}(p,s)}{\partial p} = \frac{D_1(s) - \widetilde{n}(s) p}{p(1-p)} \quad \mbox{and} \quad \mathcal{I}(p,s) = E\left\{- \frac{\partial^2 \log \widetilde{L}(p,s)}{\partial p^2} \right\} = \frac{\widetilde{n}(s)}{p(1-p)}.
\end{align*}
The maximum information for the trial is $\mathcal{I}_{\max}(p) = N / [p(1-p)]$. Under $H_0,$ the score statistic $S(p_0,s)$ has zero mean and variance $\mathcal{I}(p_0,s)$.

\cite{WangTsia87} originally proposed their designs for two sided testing to compare means of a normally distributed outcome between two treatment groups with $K$ planned analyses and equal group sizes at each analysis. For a specified design parameter $\Delta$, the decision rule rejects the null hypothesis of equal means at analysis $k$ if $|Z_k| \ge c(\alpha) (k/K)^{\Delta-0.5}$, where $c(\alpha)$ is selected to achieve type I error control at level $\alpha$. Because equally sized groups are assumed, $k/K$ is the fraction of the maximum information available at analysis $k$.  

Denote the standardized score statistic for testing $H_0: p = p_0$ by \begin{align*}
Z(s) = \frac{S(p_0,s)}{\sqrt{\mathcal{I}(p_0,s)}} = \frac{D_1(s) - \widetilde{n}(s) p_0}{\sqrt{\widetilde{n}(s) p_0 (1-p_0)}}.
\end{align*}
Our stopping criterion for one-sided testing of $H_0$ rejects at analysis time $s$ if
\begin{align*}
Z(s) \ge c_\Delta(\alpha) \left\{\frac{\mathcal{I}(p_0,s)}{\mathcal{I}_{\max}(p_0)}\right\}^{\Delta-0.5} = c_\Delta(\alpha) \left\{\frac{\widetilde{n}(s)}{N}\right\}^{\Delta-0.5},
\end{align*}
where the term in brackets is the information fraction at time $s$. Equivalently, reject if
\begin{align*}
D_1(s) \ge \widetilde{n}(s) p_0 + c_\Delta(\alpha) \sqrt{p_0 (1-p_0)} \cdot \widetilde{n}(s)^\Delta {N}^{0.5-\Delta},
\end{align*}
where the right-hand side of the inequality is the stopping boundary function 
\begin{align*}
b\left(\widetilde{n}(s)\right) = \widetilde{n}(s) p_0 + c_\Delta(\alpha) \sqrt{p_0 (1-p_0)} \cdot \widetilde{n}(s)^\Delta {N}^{0.5-\Delta}.
\end{align*}
For integer values of $\widetilde{n}(s)$, this boundary function coincides with the stopping boundary of the binary Wang-Tsiatis tests studied in \cite{MartLoga23}.

The original Wang-Tsiatis designs with $\Delta = 0.5$ and 0 corresponds to the respective Pocock \citep{Poco77} and O'Brien-Fleming designs \citep{OBriFlem79} for comparing means. For these TITE versions of Wang-Tsiatis tests for safety monitoring, the respective choices $\Delta = 0.5$ and 0 are also called Pocock and O'Brien-Fleming tests.

\subsection{Bayesian Beta-Extended Binomial Model}
\label{subsec:m_bayes}
Bayesian beta-binomial models have been employed to monitor efficacy and toxicity proportions in clinical trials \citep{ThalSimo94,GellFoll03,ChenChal06}. These models have two properties that facilitate their implementation: (1) the beta prior for the proportion is conjugate to the binomial likelihood such that the posterior distribution of the proportion is also beta-distributed, simplifying calculation of posterior probabilities; and (2) incorporation of existing knowledge into the prior distribution is straightforward.   

To retain these qualities for monitoring time-to-event toxicities, we propose an extension termed the beta-extended binomial model. This model assumes that $p$ has a $Beta(k,m)$ prior distribution and that at analysis time $s$, $D_1(s)| \widetilde{n}(s), p$ yields the extended binomial likelihood in (\ref{eq:ebin_lik}). Because this likelihood and the prior distribution are both proportional to functions of the form $p^u (1-p)^v$, this is a conjugate model with $p|D_1(s), \widetilde{n}(s) \sim Beta\big(k+D_1(s), m+\widetilde{n}(s)\big)$.

The decision rule is to reject $H_0$ at analysis time $s$ if the posterior probability that $H_0$ is false is sufficiently high, i.e. $P\big(p > p_0 | D_1(s), \widetilde{n}(s)\big) \ge c_B(\alpha)$, where $c_B(\alpha)$ is a prespecified threshold that provides a type I error rate of $\alpha$. For fixed $m$, the set of distributions $Beta(k,m), k>0$ is stochastically increasing with respect to $k$. Hence, $P\big(p > p_0 | D_1(s)=d, \widetilde{n}(s)\big) \ge c_B(\alpha)$ implies $P\big(p > p_0 | D_1(s)=d, \widetilde{n}(s)\big) \ge c_B(\alpha)$ for $D_1(s) \ge d$, showing that the decision rule can equivalently be expressed as $D_1(s) \ge b\big(\widetilde{n}(s)\big)$, where $b\big(\widetilde{n}(s)\big) = argmin_d  P\big(p > p_0 | D_1(s)=d, \widetilde{n}(s) \big) \ge c_B(\alpha)$. \cite{GellFoll03} proposed applying the beta-binomial model to monitor binary toxicities, with a boundary function and decision rule that are identical to the beta-extended binomial's at integer values of $\widetilde{n}(s)$.

Prior specification for $p$ is straightforward since its hyperparameters $k$ and $m$ may be interpreted as the respective numbers of ``prior" events and nonevents. Thus, existing data or clinician experience can guide the choices of these hyperparameters. In the absence of relevant experience, a weak prior can be specified as follows. The concentration parameter $\nu = k+m$ represents the number of prior patients. Therefore, setting $\nu$ to a small value, say 1-3, produces a low information prior. Centering the prior mean $E(p)=k/\nu$ at the null toxicity rate $p_0$ is a reasonable default, so that $p_0 = k/\nu$. These two constraints determine the hyperparameters as $k = \nu p_0$ and $m = \nu(1-p_0)$, determining the prior.

\subsection{Truncated Sequential Probability Ratio Test (SPRT)}
\label{subsec:m_sprt}
\cite{Wald45} introduced the SPRT, which repeatedly compares point null and alternative hypotheses using likelihood ratio tests. This SPRT allows early stopping both to accept and reject $H_0$ and imposes no upper limit on the sample size. To exercise vigilance in monitoring for safety issues, however, early stopping to accept $H_0$ is undesirable. Also, a clinical trial always has an upper limit on its sample size due to monetary, logistic, and feasibility constraints. To construct an SPRT that is practically useful for monitoring binary safety endpoints, \cite{Gold87} proposed a truncated SPRT that (i) only permits early stopping to reject $H_0: p = p_0$ in favor of a prespecified alternative point hypothesis $H_1: p = p_1$, where $p_1> p_0$, and (ii) truncates the test at the maximum sample size $N$ such that if $H_0$ is not rejected after $N$ patients are evaluated, $H_0$ is accepted and the procedure ends. 

We propose a truncated SPRT to evaluate $H_0$ based on (\ref{eq:ebin_lik}) to incorporate censored data when evaluating time-to-event toxicities. At analysis time $s$, the log likelihood ratio is
\begin{align*}
r(s) = \log\left(\frac{\widetilde{L}(p_1,s)}{\widetilde{L}(p_0,s)}\right) = D_1(s) \log \theta + \widetilde{n}(s) \log \left(\frac{1-p_1}{1-p_0} \right),
\end{align*}
where $\theta$ is the odds ratio of $p_1$ to $p_0$. The decision rule rejects at time $s$ if $r(s) \ge c_S(\alpha)$, or
\begin{align*}
D_1(s) \ge \frac{c_S(\alpha) - \widetilde{n}(s) \{\log (1-p_1) - \log (1-p_0)\}}{\log \theta},
\end{align*}
where the critical value $c_S(\alpha)$ achieves a false positive rate of $\alpha$. The boundary function is 
\begin{align*}
b\left(\widetilde{n}(s)\right) = \frac{c_S(\alpha) - \widetilde{n}(s) \{\log (1-p_1) - \log (1-p_0)\}}{\log \theta} .
\end{align*}
The proposed truncated SPRT's stopping boundary function matches Goldman's truncated SPRT for binary toxicities at integer values of $\widetilde{n}(s)$.

\subsection{Constructing Stopping Boundaries with Type I Error Control}
\label{subsec:m_alpha}
We intend for these methods to achieve a type I error rate of $\alpha$, which requires tuning of the critical values $c_\Delta(\alpha)$, $c_B(\alpha)$, and $c_S(\alpha)$ for the Wang-Tsiatis tests, Bayesian beta-extended binomial models, and truncated SPRTs, respectively. Because the enrollment pattern and timing of events are unpredictable for a given trial and the construction of the extended binomial likelihood entails approximation of the true likelihood, an exact, direct calculation of the type I error rate for a given boundary does not appear to be possible. But, the boundary functions for these methods can be viewed as extensions of the boundaries for analogous binary data monitoring techniques to non-integer values of  $\widetilde{n}(s)$. Therefore, we propose that the necessary critical values be computed under a binary toxicity data monitoring process, i.e. each patient's toxicity outcome is treated as a binary occurrence and only evaluated when they have fully completed follow-up through the evaluation window $[0,\tau]$. Online Supplemental Appendix B of \cite{MartLoga23} describes how to calibrate the critical value for each method to achieve the type I error rate closest to and not exceeding $\alpha$ under binary toxicity monitoring. Though we do not expect that this calibration will provide strict type I error control for the TITE-based monitoring approaches, our simulation studies show that the type I error rates are maintained near $\alpha$ in a variety of settings, including ones where the event distribution deviates greatly from a uniform distribution.

\subsection{Design Considerations}
\label{subsec:m_design}
Each TITE stopping rule approach needs some design parameter(s) to be prespecified. The null cumulative incidence $p_0$ of toxicity at time $\tau$ is required for all methods. In addition, the Wang-Tsiatis test requires a choice of $\Delta$, the beta-extended binomial model needs the number of prior patients $\nu$, and the truncated SPRT needs a choice of excessive alternative cumulative incidence, $p_1$. These parameters can be tuned to modify the boundary's shape as a function of $\widetilde{n}(s)$, in turn adjusting its performance metrics. When designing safety stopping rules for a new trial, operating characteristics including the type I error rate, power to detect excess safety risk, and expected number of patients suffering toxicity should be evaluated and the rules adjusted as necessary to meet the trial's requirements for vigilance.

\section{Simulation Study}
\subsection{Overview}
\label{subsec:s_overview}
To study the proposed methods, we consider their usage for monitoring cohorts of size $N = 50$ and 200, the respective typical sizes of study arms in phase 2 and 3 trials. The methods' stopping boundaries and operating characteristics are examined for each cohort size.

\subsection{Stopping Boundary Shapes}
\label{subsec:s_bdry}
The stopping boundaries of the safety monitoring approaches are considered for the scenarios where the evaluation window is $\tau=30$ days and the expected toxicity rate is $p_0 = 0.1$ and 0.25. Six stopping rules are considered: Pocock test (POC); O'Brien-Fleming test (OBF); truncated SPRTs with alternative rates $p_1$'s that a fixed sample, exact binomial test can detect with $65\%$ power (SPL) and $95\%$ power (SPH); and beta-extended binomial models with prior means $E(p) = p_0$ and weakly and strongly informative priors determined by numbers of prior patients $\nu = 1$ (BW) and $\nu = 0.25N$ (BS), the latter constituting 25\% of the cohort size. The stopping boundary of each method was calibrated to have a 5\% nominal type I error rate under a binary toxicity monitoring process.
 
\begin{figure}
\caption{Stopping Boundaries from Safety Monitoring Methods. Panels A and C show the stopping boundaries in a 50 patient cohort, while panels B and D displays boundaries for a 200 patient cohort. Boundaries in the top row were constructed for a null toxicity rate of $p_0=0.1$, while the bottom row considers $p_0=0.25$. Boundaries are displayed as the number of events that would trigger a rejection of $H_0: p=p_0$ versus the effective sample size computed among enrolled patients. Methods considered include Pocock test (POC); O'Brien-Fleming test (OBF); truncated SPRT, low $p_1$ (SPL); truncated SPRT, high $p_1$ (SPH); beta-extended binomial model, weak prior (BW); and beta-extended binomial model, strong prior (BS).}
\begin{center}
\includegraphics[width=0.95\textwidth]{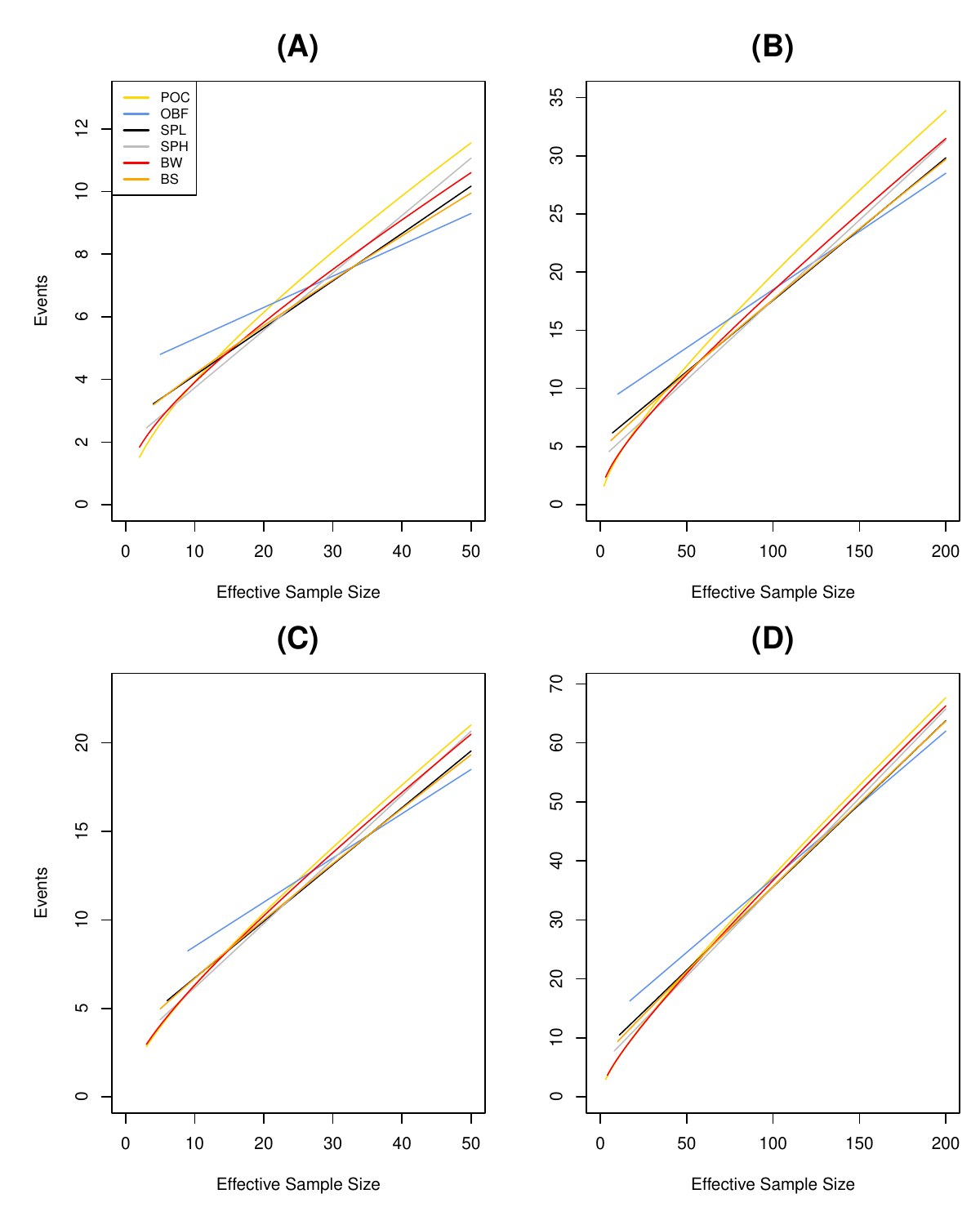}
\end{center}
\label{stopbounds}
\end{figure}

Figure \ref{stopbounds} displays the stopping boundaries for cohort sizes of $N=50$ and $200$ and $p_0=0.1$ and 0.25, plotted over the range of effective sample sizes $[0,N]$. The Pocock boundaries are the most aggressive early but most lenient during the middle and late portions of the trial. The O'Brien-Fleming test, on the other hand, has the most conservative boundaries early and most aggressive ones late in the study. Among the beta-extended binomial models, the weak prior specification yields boundaries that are more strict early and more permissive late in the trial compared to the strong prior model. The truncated SPRTs exhibit a similar pattern with the higher choice of $p_1$ (SPH) imposing more aggressive early stopping and more lenient late stopping criteria compared to the low $p_1$ version (SPL). The SPL and BW methods have similar boundaries throughout most of the monitoring period, as do SPH and BS. The variety of shapes of these boundaries illustrate that a trade-off of safety vigilance is required across analyses in order to control type I error across repeated evaluations.

\subsection{Operating Characteristics}
\label{sec:s_oc}
We next examine the operating characteristics of these monitoring methods. Settings considered included cohort sizes of $N=50$ and $200$, null event proportions of $p_0 = 0.1$ and $0.25$, and evaluation window lengths of $\tau=30$ and $100$ days. Uniform enrollment over a 2 year period was assumed. Scenarios both with and without competing risks were considered; for the former, the cumulative incidence of the competing risk at time $\tau$ was 0.3. For scenarios with competing risks, the cumulative incidence function for the toxicity event of interest is $F_1(t) = pt/\tau$ and the cumulative incidence function for all competing risks combined is $F_2(t) = 0.3t/\tau$ for $t \in [0,\tau/(p+0.3)]$, such that at time $\tau$ the cumulative incidences of the event of interest and competing risks are $p$ and $0.3$, respectively. In settings without competing risks, the cumulative incidence function of the toxicity event of interest is $F_1(t) = pt/\tau$, $t \in [0,\tau/p]$. The type I error rate, power, and expected number of toxicities were evaluated for each scenario using 10,000 simulated datasets. Each method was calibrated to have a 5\% nominal type I error rate under a binary toxicity monitoring process.

Figures 2 and 3 display empirical estimates of the power and expected toxicities scenarios for $N=50$ and 200, respectively, with $\tau=30$ and where competing risks are at play. Results for configurations with $\tau=100$ and without competing risks were similar and are omitted. The type I error rates for all approaches are controlled at 5\% or less. All methods except for the Pocock test have type I error rates between 4\% and 5\%; the Pocock test generally has the lowest rates among all methods, with 3-4\% type I error rates in scenarios with $N=200$. When $p>p_0$, the O'Brien-Fleming test has the highest power, followed by BS and SPL. The Pocock test has the lowest power among the methods considered. SPH has the lowest number of expected toxicities, followed by BS and SPL. The O'Brien-Fleming and Pocock tests yield the highest numbers of toxicities on average. The SPL test produces slightly higher power and higher numbers of toxicities compared to SPH. Among the beta-extended binomial models, BW has lower power and generally has higher expected toxicities than BS.

\begin{figure}
\caption{Operating Characteristics from Safety Monitoring Methods for $N=50$. Panels A-B show empirical estimates of power and expected number of toxicities when $p_0=0.1$ and panels C-D display these characteristics when $p_0=0.25$. Methods considered include Pocock test (POC); O'Brien-Fleming test (OBF); truncated SPRT, low $p_1$ (SPL); truncated SPRT, high $p_1$ (SPH); beta-extended binomial model, weak prior (BW); and beta-extended binomial model, strong prior (BS).}
\begin{center}
\includegraphics[width=0.95\textwidth]{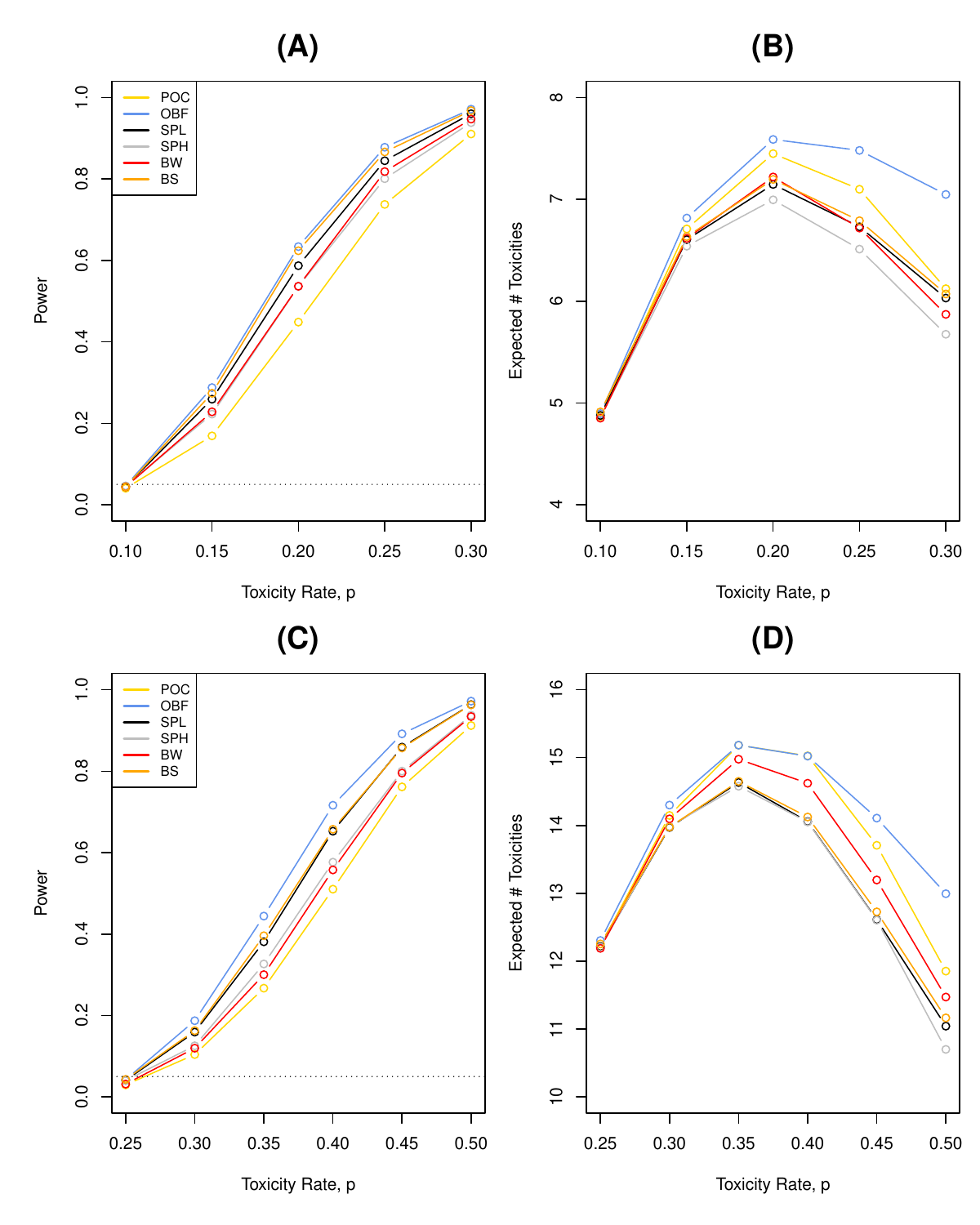}
\end{center}
\label{fig:opchar_tite_n50}
\end{figure}

\begin{figure}
\centering
\caption{Operating Characteristics from Safety Monitoring Methods for $N=200$. Panels A-B show empirical estimates of power and expected number of toxicities when $p_0=0.1$ and panels C-D display these characteristics when $p_0=0.25$. Methods considered include Pocock test (POC); O'Brien-Fleming test (OBF); truncated SPRT, low $p_1$ (SPL); truncated SPRT, high $p_1$ (SPH); beta-extended binomial model, weak prior (BW); and beta-extended binomial model, strong prior (BS).}
\includegraphics[width=0.95\textwidth]{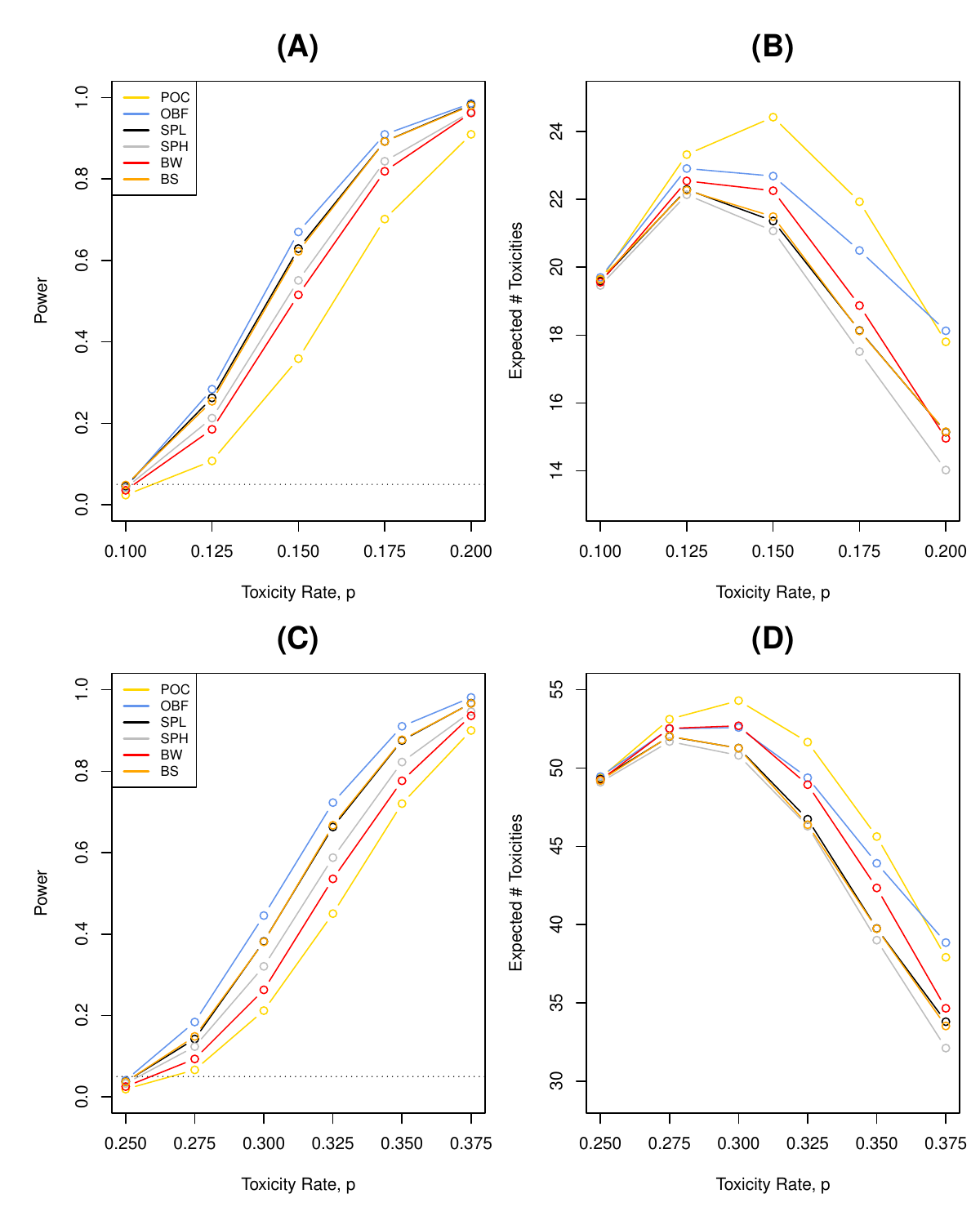}
\label{fig:opchar_tite_n200}
\end{figure}

The alpha spending functions were estimated as well for each method for $N=50$ and 200, $p_0=0.1$, and $\tau=30$ in the presence of competing risks; results are displayed in Figures 4-5. We attribute the inferior power and high expected toxicities of the Pocock test to its highly aggressive stopping criteria early in trials. During the monitoring of the first 10\% of patients, this test spends 2.5\% or more of the total type I error, leading to highly conservative evaluations in the remainder of the study that yield diminished power. The BW method also spends alpha quickly in the early portion of the study, causing it to have lower power and higher toxicities than SPH and BS.

\begin{figure}
\centering
	\caption{Empirical Cumulative Type I Error Rates from Safety Monitoring Methods for $N=50$, $p_0=0.1$, $\tau=30$. Methods considered include Pocock test (POC); O'Brien-Fleming test (OBF); truncated SPRT, low $p_1$ (SPL); truncated SPRT, high $p_1$ (SPH); beta-extended binomial model, weak prior (BW); and beta-extended binomial model, strong prior (BS).}
	\includegraphics*[width=0.9\textwidth]{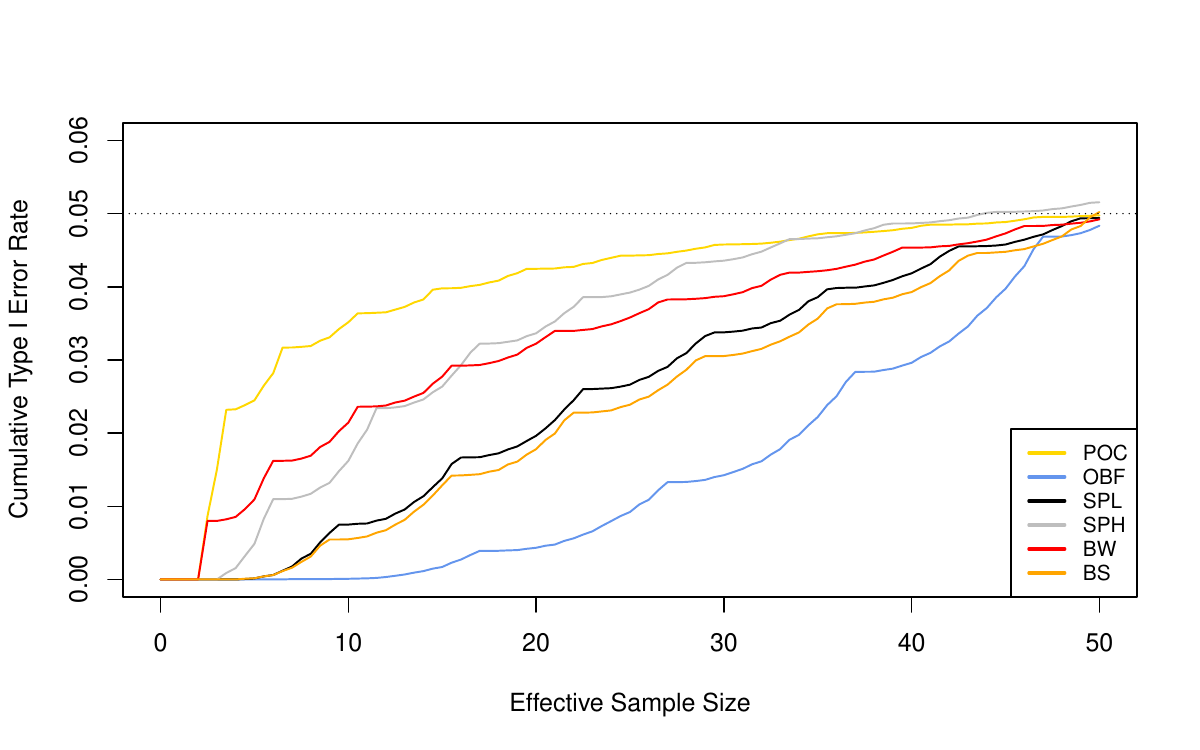}
\label{fig:opchar_surv_n50}
\end{figure}

\begin{figure}
\centering
	\caption{Empirical Cumulative Type I Error Rates from Safety Monitoring Methods for $N=200$, $p_0=0.1$, $\tau=30$. Methods considered include Pocock test (POC); O'Brien-Fleming test (OBF); truncated SPRT, low $p_1$ (SPL); truncated SPRT, high $p_1$ (SPH); beta-extended binomial model, weak prior (BW); and beta-extended binomial model, strong prior (BS).}
	\includegraphics*[width=0.9\textwidth]{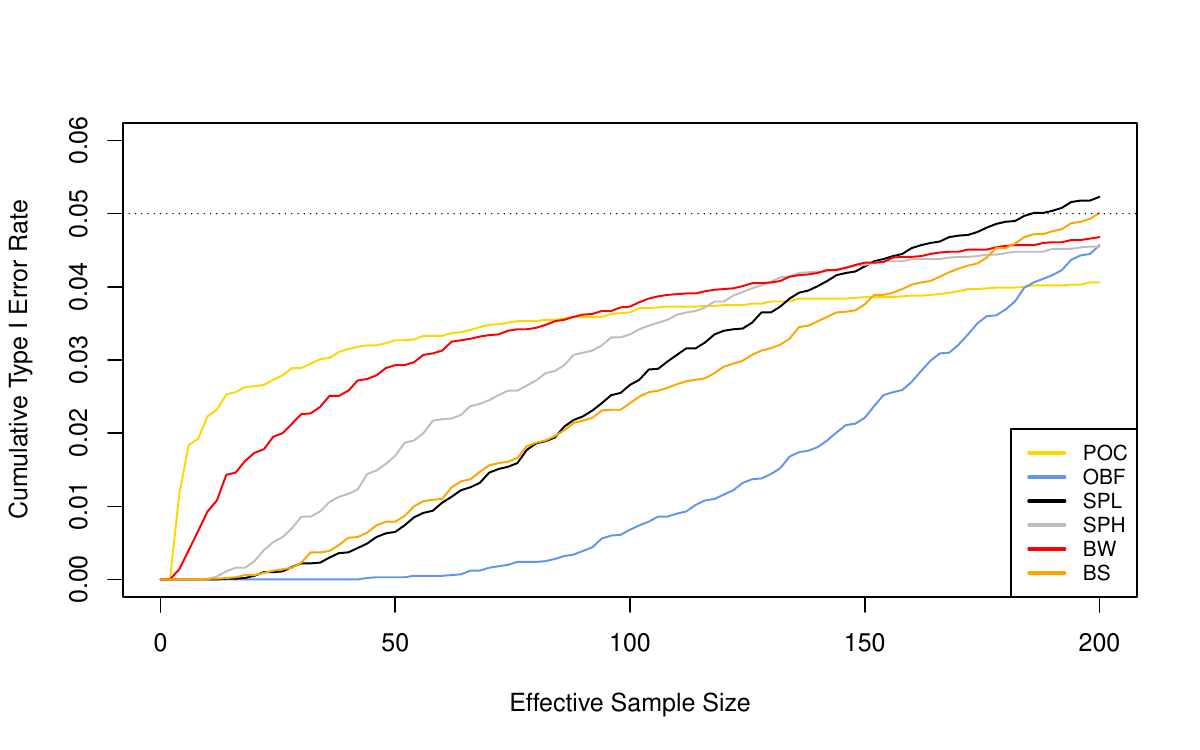}
\label{fig:opchar_surv_n200}
\end{figure}

\subsection{Comparison of TITE Methods to Binary Safety Monitoring Rules}
We also compared the performance of the proposed, TITE safety monitoring approaches to existing binary toxicity data methods. The binary monitoring rules require a patient to complete all follow-up through their assessment window $[0,\tau]$ to be evaluated. Because the TITE-based methods utilize partial follow-up information on patients who are event-free but still being followed in $[0,\tau]$, they may produce favorable operating characteristics compared to those of binary monitoring strategies. To evaluate this hypothesis, analogous binary stopping rules of the 6 proposed TITE-based methods were constructed with nominal 5\% type I error rates: binary Pocock and O'Brien-Fleming tests; Bayesian beta-binomial models with weak and strong priors determined by respective concentration parameter values of $1$ and $0.25N$; and binary truncated SPRTs with alternative rate $p_1$'s that a fixed sample binomial test of $p$ can detect with $65\%$ and $95\%$ power. Type I error rates, power, and expected toxicities were computed directly for the binary monitoring methods as described in Online Appendix A of \cite{MartLoga23}, assuming uniform accrual over 2 years. 

\begin{figure}
\caption{Comparison of Operating Characteristics for Proposed TITE-based and Binary Stopping Rule Methods. Panels A and C display respective plots under $\tau=30$ and $100$ of the power levels attained for corresponding TITE and binary rules within each method when $N=50$ and $p_0=0.1$. Panels B and D show respective plots under $\tau=30$ and $100$ of the ratio of expected numbers of toxicities for corresponding TITE-based and binary rules within each method when $N=50$ and $p_0=0.1$. Methods considered include Pocock test (POC); O'Brien-Fleming test (OBF); truncated SPRT, low $p_1$ (SPL); truncated SPRT, high $p_1$ (SPH); Bayesian model, weak prior (BW); and Bayesian model, strong prior (BS).}
\begin{center}
\includegraphics*[width=0.95\textwidth]{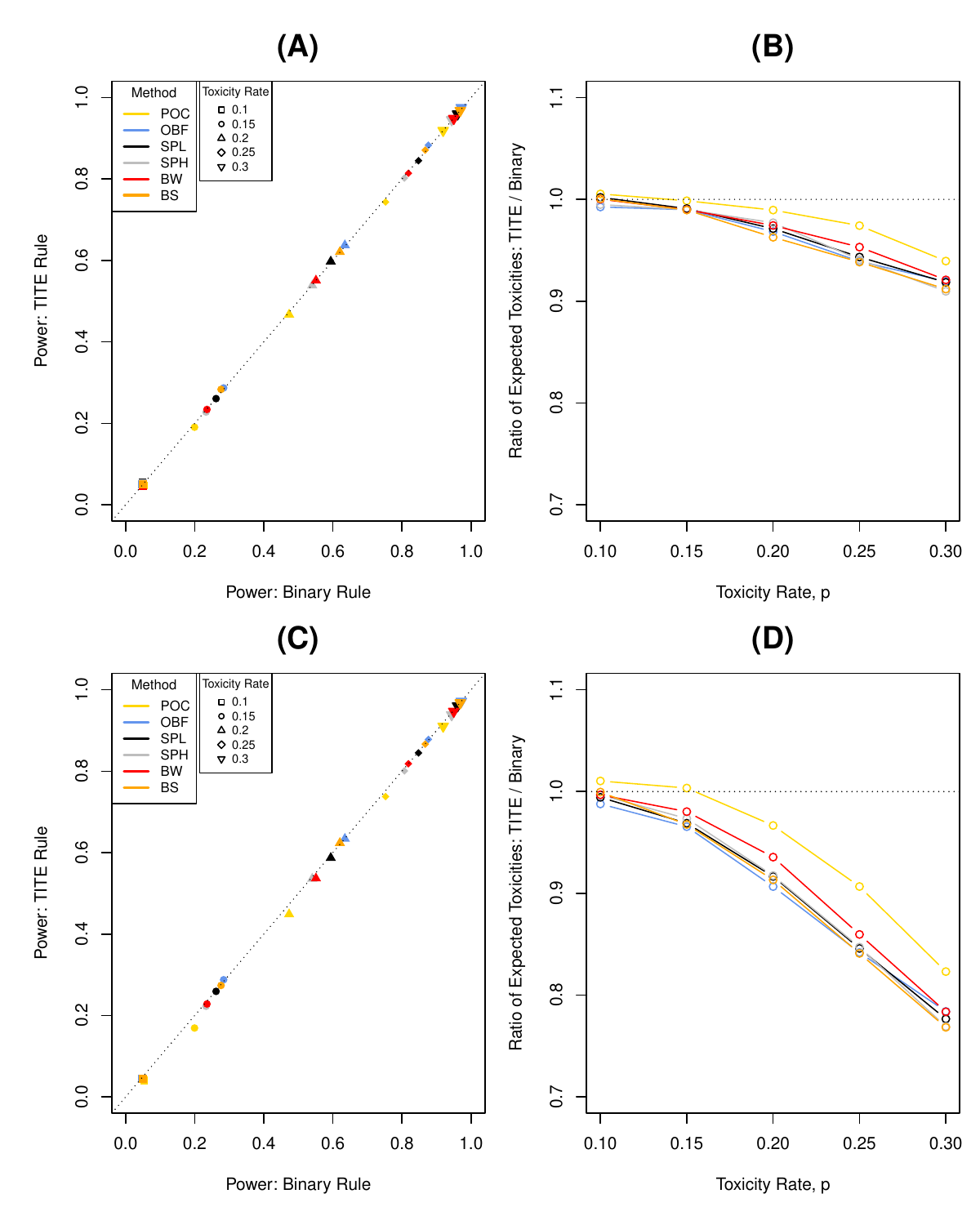}
\end{center}
\label{fig:titebin_compare}
\end{figure}

The power and expected toxicities are contrasted between the TITE and binary stopping rules in 
Figure \ref{fig:titebin_compare} for scenarios with $N=50$, $p_0=0.1$, $\tau=30$ and $100$, and with competing risks at play. The power levels of TITE and binary monitoring rules are similar within each of the 6 testing methods and do not differ markedly between scenarios with $\tau=30$ and $100$. When $p>0.10$, all approaches except for the Pocock test produce lower numbers of toxicities for the TITE-based rule compared to its binary version. For excessive toxicity rates of $0.25$ or more, the TITE approach yields 5-8\% fewer toxicities when $\tau=30$ and a reduction of 15-22\% in toxicities when $\tau=100$. As $\tau$ increases, more patients tend to have partial follow-up for the toxicity at an interim analysis, providing more partial information that the TITE-based methods can leverage but the binary methods ignore. We also conducted comparisons for scenarios with $p_0=0.25$, $N=200$, and without competing risks, which observed similar advantages of the TITE-based stopping rule methods over binary ones (results not shown). This comparison shows that TITE approaches offer substantial reductions in the numbers of patients suffering toxicity without compromising the power to identify safety issues. 

\subsection{Comparison of TITE Methods to Censored Exponential Safety Monitoring Rules}
For time-to-event toxicities without competing risks, \cite{MartLian25} previously developed safety monitoring approaches that treat the toxicity data as survival outcomes under a censored exponential model. Under this assumption, the exponential event rate $\lambda = -\log(1-p) / \tau$ is evaluated and the null hypothesis may be equivalently expressed as $H_0^{CE}: \lambda = \lambda_0$, where $\lambda_0 = -\log(1-p_0) / \tau$. These methods are effective in controlling the type I error rate when the data truly follow an exponential distribution, but may be subject to type I error inflation or diminished power when the true data generating distribution deviates from an exponential. Because the weights $w_j(s)$ used in the extended binomial likelihood are constructed assuming a uniform event distribution on $[0,\tau]$, the TITE-based stopping rules are expected to provide good type I error control when event data are generated from a uniform distribution but may not necessarily control type I error well for other event distributions. Therefore, we conducted another set of simulations both to assess the robustness of the TITE monitoring approaches with regard to type I error rate control and to contrast these rates between the TITE and censored exponential based stopping rules. 

The following censored exponential stopping rule analogues of the 6 proposed TITE-based methods were evaluated with nominal 5\% type I error rates: censored exponential Pocock and O'Brien-Fleming tests; Bayesian gamma-Poisson models with weak and strong priors determined by prior mean $E(\lambda) = \lambda_0$ and prior rate parameter values of $\tau$ and $0.25N \tau$, respectively, the prior rates representing ``total prior exposure time"; and truncated SPRTs with alternative rates $p_1$'s that a fixed sample, exact test of the Poisson mean $\lambda \cdot N \tau$ can detect with $65\%$ and $95\%$ power. Scenarios with $N=50$ and 200, $p_0=0.1$ and 0.25, $\tau=30$, and no competing risks were examined. Toxicity data were generated to produce event probabilities of $p_0$ at time $\tau$ under uniform, exponential, Weibull - shape parameter = 0.5, and Weibull - shape 2.0 distributions. For each scenario, 10,000 datasets were simulated and analyzed.   

\begin{figure}
\caption{Comparison of Operating Characteristics for Proposed Survival and Binary Stopping Rule Methods. Panels A and C show respective scatter plots under $\tau=30$ and $100$ of the power levels attained for corresponding survival and binary rules within each method when $n=50$ and $p_0=0.1$. Panels B and D show respective series plots under $\tau=30$ and $100$ of the ratio of expected numbers of events for corresponding survival and binary rules within each method when $n=50$ and $p_0=0.1$, respectively. Methods considered include Pocock test (POC); Wang-Tsiatis test with $\Delta=0.25$ (WT=0.25); O'Brien-Fleming test (OBF); truncated SPRT, low $p_1$ (SPL); truncated SPRT, high $p_1$ (SPH); MaxSPRT (MSP); Bayesian model, weak prior (BW); and Bayesian model, strong prior (BS).}
\begin{center}
\includegraphics*[width=0.95\textwidth]{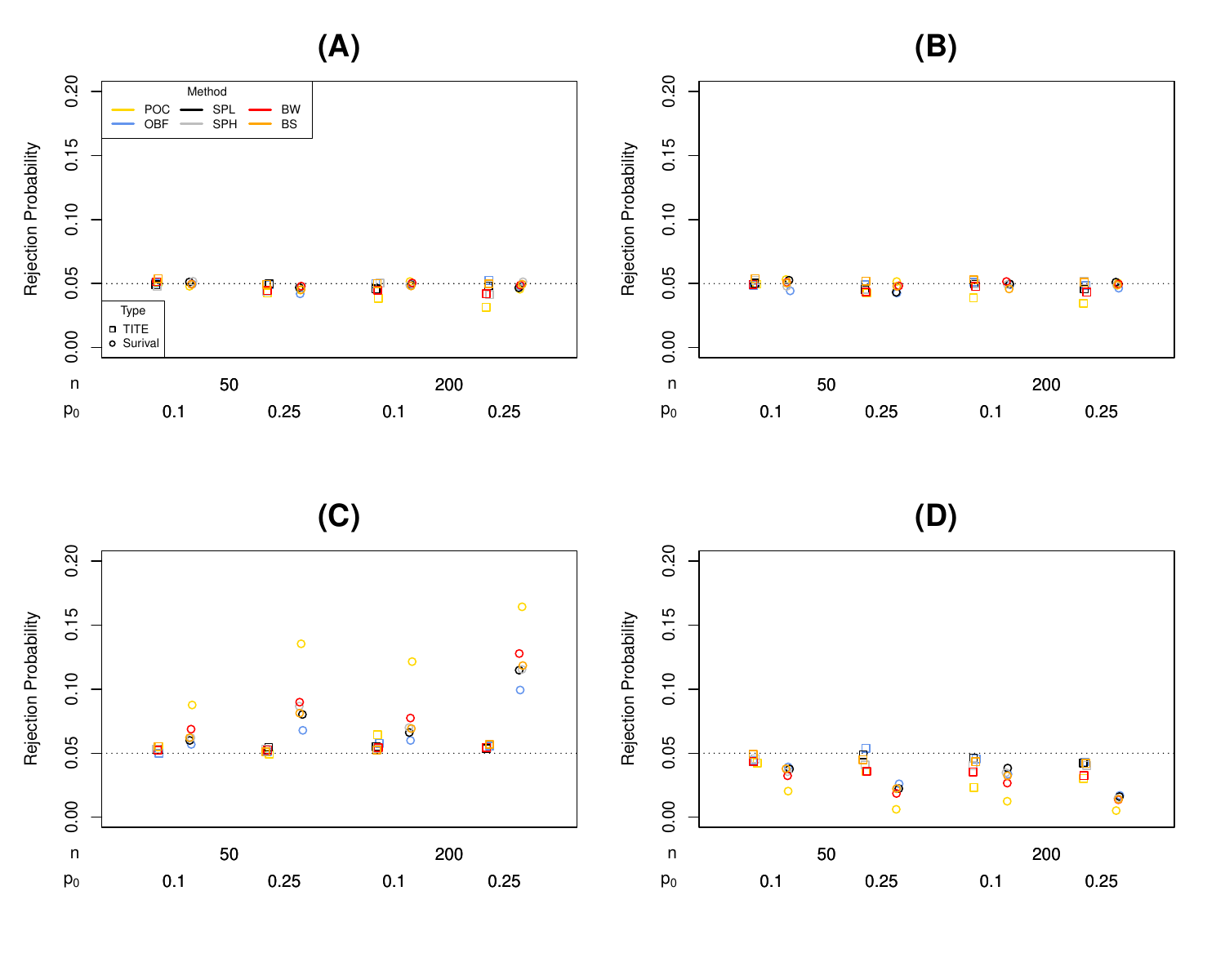}    
\end{center}
\label{fig:titesurv_compare}
\end{figure}

The empirical type I error rates are shown in Figure \ref{fig:titesurv_compare}. Under uniform and exponential event distributions, the type I error rates of both the TITE and censored exponential approaches are generally well controlled near 5\%, with the TITE Pocock test showing conservative rates of 4\% or lower. When the data are generated from an Weibull - shape 0.5 distribution, the TITE methods have type I error rates in the 5-7\% range across scenarios, while the censored exponential stopping rules show greater inflation with most rates exceeding the TITE techniques' and some rates exceeding 10\%. Conversely, a Weibull - shape 2.0 event distribution yields the type I error rates for the TITE rules in the 2-5\% range, which are generally closer to 5\% than the censored exponential rules rates that lie in the range 1-4\%. Overall, the TITE stopping rules are fairly robust to event distributions that differ greatly from a uniform one, while the censored exponential rules are sensitive to deviations from its assumption of exponentially distributed toxicities. 

\section{Example}
The application of the proposed safety monitoring methods are illustrated using data from the BMT CTN 0601 trial, a phase II, multicenter, single arm trial that assessed the efficacy and safety of bone marrow transplant as treatment for pediatric patients with severe sickle cell disease. The target enrollment was 30 patients. To facilitate the use of TITE-based stopping rules for trial design, our R package "stoppingrule" has functions for constructing and evaluating the operating characteristics of these methods. These functions are demonstrated through a redesign of the safety monitoring scheme of BMT CTN 0601. 

The enrollment target for the trial was 30 patients over a 4 year accrual period. Since the safety profile of bone marrow transplant for sickle cell disease patients was not well understood, the trial installed stopping rules to monitor three safety events for excess risk: early mortality, graft rejection, and severe graft-versus-host disease (GVHD). Based on prior research, it was expected that up to 15\% of patients would die, 15\% would develop severe GVHD, and 20\% would have graft rejection by Day 100 post-transplant. The stopping rules of the original trial were constructed using truncated SPRTs for binary toxicity outcomes, each having an 8\% type I error rate; the early mortality and severe GVHD rules had identical stopping criteria since the same null event rates of 15\% and type I error rates were used.

For our redesign of the safety monitoring scheme, three candidate TITE-based methods with nominal type I error rates of 5\% are considered for trial planning: O'Brien-Fleming test; truncated SPRT, high $p_1$; and beta-extended binomial model, $0.25N = 7.5$ prior patients. The beta-extended binomial stopping criteria for early mortality and severe GVHD can be constructed with hyperparameters $7.5 \cdot (0.15, 0.85)$ using the R function \textbf{calc.rule.tite}:

\begin{verbatim}
bayes_rule = calc.rule.tite(n=50, p0=0.15, alpha=0.05, type="Bayesian",
                            param=7.5*c(0.15, 0.85))
\end{verbatim}

The output of this function, \textbf{bayes\underline{{ }{ }}rule}, is a list comprised of the stopping rule's design specifications, stopping boundary, and boundary parameter $c_B(\alpha)$. The stopping rule can be tabulated succinctly using the \textbf{table.rule.tite(bayes\underline{{ }{ }}rule)} command:

\begin{verbatim}
Effective Sample Size   Reject Bdry
     4.00 -  6.99            4
     7.00 - 11.27            5
    11.28 - 15.73            6
    15.74 - 20.34            7
    20.35 - 25.08            8
    25.09 - 29.92            9
    29.93 - 30.00           10
\end{verbatim}

This rule halts the trial if four events occur when the effective sample size is 4.00 - 6.99, 5 events occur when this size is less than or equal to 11.27,  6 events happen among the first 15.73 ``patients", and so on.

Calling the \textbf{plot()} and \textbf{lines()} functions with a \textbf{calc.rule.tite} object as the input will construct a plot of the stopping rule boundary and add the boundary to an existing plot, respectively. Figure 8 displays the stopping boundaries of the three candidate TITE-based monitoring rules for severe GVHD and graft rejection, which were drawn using these functions. The truncated SPRTs have the most aggressive boundaries early and most lenient boundaries late in the trial. The O'Brien-Fleming tests exhibits the opposite pattern, being most conservative with stopping early and most strict late in the trial. The beta-extended binomial rules strikes a balance between the SPRTs and O'Brien-Fleming tests with regard to early vs. late aggressiveness in stopping criteria. The trial enrolled 30 patients, of whom 29 received a transplant and were evaluable for the stopping rules. Among transplanted patients, none died, five developed severe GVHD, and three had a graft rejection by Day 100 post-transplant; no stopping criteria were flagged by these events in the original trial. Figure 8 displays the cumulative severe GVHD and graft rejection events computed daily during the course of the trial, showing that no safety issues would have been flagged for any of the candidate rules for mortality, severe GVHD, and graft rejection. 

\begin{figure}
\begin{center}
\caption{Stopping Rule Boundaries and Observed Toxicity Data in BMT CTN 0601 Population. Panel A shows the severe GVHD stopping boundaries and events over the course of the trial, while panel B displays graft rejection stopping rules and events.}
\end{center}
\includegraphics[width=1.00\textwidth]{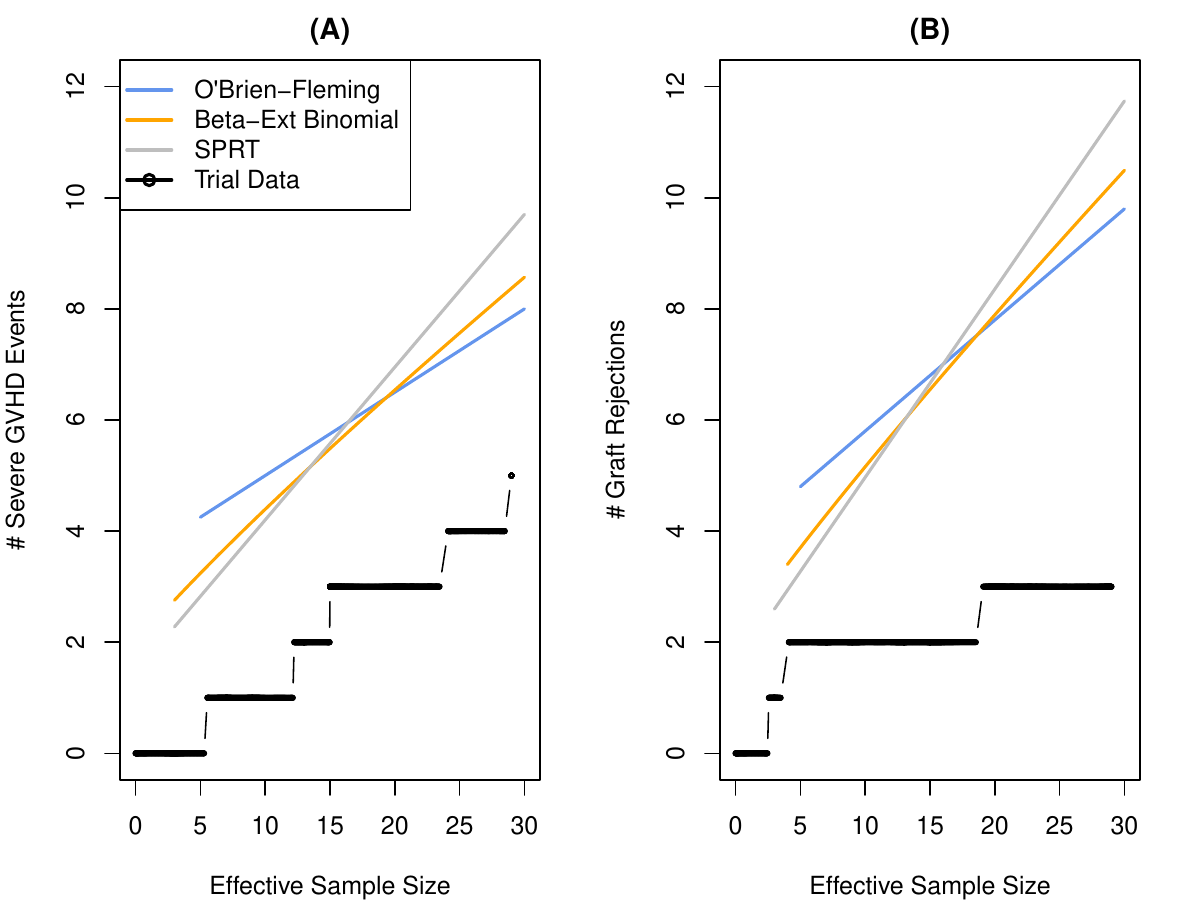}
\label{fig:example}
\end{figure}

At the trial design stage, an examination of the performance metrics of candidate stopping rules is essential for selecting the most appropriate safety monitoring regime. The function \textbf{OC.rule.tite()} can perform this evaluation for a given stopping rule. For the beta-extended binomial rule with a 4 year (1,460 day) enrollment period, the function call

\begin{verbatim}
OC.rule.tite(rule=bayes_rule, ps=seq(0.15,0.35,0.05), ps.compt=rep(0.10,5),
             tau=100, MC=10000, A=1460)
\end{verbatim}

\noindent
runs 10,000 Monte Carlo simulations to estimate the performance metrics at $p = 0.15$, 0.20, 0.25, 0.30, and 0.35 for this rule, assuming the cumulative incidence of the competing risk, death without prior severe GVHD, is 0.1 at Day 100, the severe GVHD event times follow a uniform distribution, and enrollment times are uniformly distributed. Operating characteristics evaluated include the rejection probability, expected number of events, expected number of patients enrolled, and expected study duration. Computing these for the Bayesian beta-extended binomial model, O'Brien-Fleming test (OBF), and truncated SPRT yields the following output: 

\begin{verbatim}
type    p     p.compt  Reject Prob  E(Events)  E(Enrolled)  E(Calendar time)
OBF    0.15     0.1      0.0481      4.4441     29.6118        1534.7123
OBF    0.20     0.1      0.1708      5.6454     28.6136        1470.6995
OBF    0.25     0.1      0.3873      6.5298     26.5861        1344.3839
OBF    0.30     0.1      0.6224      6.9160     23.6450        1171.8715
OBF    0.35     0.1      0.8165      6.9434     20.5543         997.7528
Bayes  0.15     0.1      0.0509      4.4196     29.3616        1522.9662
Bayes  0.20     0.1      0.1696      5.5569     28.0121        1442.4418
Bayes  0.25     0.1      0.3875      6.3015     25.3150        1284.3967
Bayes  0.30     0.1      0.6028      6.5149     22.3018        1111.0220
Bayes  0.35     0.1      0.7934      6.3591     18.8757         922.1758
SPRT   0.15     0.1      0.0471      4.3487     29.1536        1513.1554
SPRT   0.20     0.1      0.1561      5.4859     27.4231        1416.3314
SPRT   0.25     0.1      0.3295      6.1599     24.8795        1270.8607
SPRT   0.30     0.1      0.5417      6.3288     21.6262        1087.4410
SPRT   0.35     0.1      0.7473      6.0377     17.8354         880.9108
\end{verbatim}

The O'Brien-Fleming test achieves the highest power, number of toxicities, and study duration, whereas the SPRT has the lowest levels of these metrics. At an excessive acute GVHD rate of 35\%, the SPRT provides 74.7\% power, while the Bayesian approach and O'Brien-Fleming tests attain respective power levels of 79.3\% and 81.7\%. Because a high chance detecting excess risk likely a requirement of trial investigators, the latter two approaches are likely preferable to the SPRT. The Bayesian method yields 0.6 fewer toxicities and 75 fewer days in study duration on average than O'Brien-Fleming under this excessive rate. For monitoring severe GVHD in this example, the beta-extended binomial model may be preferred among the candidate rules in order to minimize the time to signal and number of patients suffering toxicity while maintaining adequate power.

\section{Discussion}
This manuscript proposes the TITE-Safety approach to monitor time-to-event toxicity endpoints while properly accounting for competing risks by adapting existing statistical methods for binary safety outcomes. Rigorous simulations examined the operating characteristics of these methods. These techniques were illustrated in a redesign of the safety stopping rules for the BMT CTN 0601 trial. Furthermore, our R package ``stoppingrule" has functions to construct stopping boundaries and assess operating characteristics of these rules, facilitating the design of a suitable monitoring scheme for a given trial.

TITE-based stopping rules with aggressive stopping criteria early in a trial tend to yield lower power and lower expected numbers of toxicities and patients enrolled than rules with more lenient early stopping boundaries. This mirrors trends seen previously for the binary and Poisson process-based safety monitoring methods \citep{MartLoga23,MartLian25} as well as group sequential tests of efficacy endpoints \citep{WangTsia87,JennTurn99}. Among the rules studied via simulation, the ones with the lowest numbers of toxicities were the truncated SPRTs and the beta-extended binomial model with a strong prior. Because the goal of safety monitoring is to minimize the risks posed to patients by flagging issues quickly, methods with the lowest numbers of toxicities and patients enrolled may be favored over those with higher numbers for these metrics and higher power, such as the O'Brien-Fleming test.

When comparing the TITE-based methods to binary monitoring techniques, the TITE stopping rules allowed 5-22\% fewer toxicities across scenarios, with greater reductions seen for $\tau=100$ compared to $\tau=30$. More patients will have incomplete follow-up for evaluation as $\tau$ and as the accrual rate increase, allowing the TITE rules to enjoy a greater reduction in numbers of toxicities and time to signal. To realize these benefits fully, (i) new safety events need to be reported quickly after their occurrences and (ii) the stopping rules must be checked promptly after new events are reported. Adequate trial staffing, a robust safety data capture system, and statistical programming of automated stopping criteria checks are required for the TITE monitoring approaches to be most effective.

The robustness of the TITE-based stopping rules to the true event time distribution was assessed and contrasted to methods based on a Poisson process assumption for event counts. The TITE methods were much more robust with regard to type I error control and power levels than Poisson based techniques across event distributions with mostly early events, mostly late events, and evenly distributed event times across the evaluation window. Since not only the toxicity rate but also the timing of toxicities are uncertain when designing the trial, this reliability of the TITE approaches across a variety of toxicity distributions makes them especially valuable for trials with time-to-event safety outcomes.

Trials often plan to monitor multiple toxicities for excess risk. The common practice is to construct separate stopping rules for each toxicity; the original design of the BMT CTN 0601 safety monitoring scheme did this and we also constructed separate stopping criteria for our redesign. But, this practice inflates the familywise error rate of the monitoring scheme and ignores information about associations between toxicities. A strategy that jointly evaluates the risks of toxicities can both control this error rate and use information about their correlations to more efficiently assess safety. We are unaware of any literature on monitoring of multiple safety outcomes, though. We plan to develop joint safety monitoring procedures for multiple toxicities in the future.  

For confirmatory trials, it may be desirable to compare toxicity risks between study arms, permitting an unbiased comparison of the safety profiles of the investigational and control therapies. Sparse literature is available on comparisons of toxicity rates between arms, though. \cite{ZhuYao16} proposed Bayesian beta-binomial models for comparing binary toxicity rates and Gamma-Poisson models for time-to-event toxicities under a Poisson event process assumption, but provided no methods to ensure false positive rate control. \cite{BollWhit00} proposed a group sequential test for contrasting toxicity rates between arms which relies on large sample properties for type I error control and, thus, may not be valid with small sample sizes or frequent evaluations. We will develop sequential methods for comparing toxicity rates with accurate type I error control for both small and large cohorts.

\section*{Acknowledgements}
Support for this study was provided by grant \#U10HL069294 to the Blood and Marrow Transplant Clinical Trials Network (BMT CTN) from the National Heart, Lung, and Blood Institute and the National Cancer Institute, along with funding by the National Marrow Donor Program, the Sickle Cell Disease Clinical Research Network, and the National Center on Minority Health and Health Disparities. The authors thank the BMT CTN for permitting use of the BMT CTN 0601 trial data. The content is solely the responsibility of the authors and does not necessarily represent the official views of the above mentioned parties.

\bibliographystyle{apalike}
\bibliography{refs}

\end{document}